\documentclass[12pt]{article}
\usepackage{graphicx}

\usepackage{amstext}
\usepackage{amssymb}
\usepackage{amsmath}
\usepackage{epsfig}
\usepackage{float}
\usepackage{subfigure}

\begin{document}
\title{\bf The flow generated by an active olfactory system of the red swamp crayfish}
\author{P. Denissenko$^1$, S. Lukaschuk$^1$, T. Breithaupt$^2$}

\maketitle

\noindent{$^1$ Fluid Dynamics Laboratory, University of Hull, HU6
7RX, UK}

\noindent{$^2$ Department of Biology, University of Hull, HU6 7RX,
UK}

\vspace{10mm} \centerline{\bf Summary} \vspace{5mm}

{\bf Crayfish are nocturnal animals that mainly rely on their
chemoreceptors to locate food. On a crayfish scale chemical stimuli
received from a distant source are dispersed by an ambient flow
rather than molecular diffusion. When the flow is weak or absent,
food search can be facilitated by currents generated by the animal
itself. Crayfish employ their anterior fan organs to produce a
variety of flow patterns. Here we study the flow generated by {\it
Procambarus clarkii} in response to odour stimulation. We found that
while searching for food the crayfish generates one or two outward
jets. These jets create an inflow which draws odour to the
crayfish's anterior chemoreceptors. We quantified velocity fields in
the inflow region using Particle Image Velocimetry. The results show
that the inflow velocity decreases proportionally to the inverse
distance from the animal so that it takes about a minute for an
odour plume to reach the animal's chemoreceptors from a distance of
10 cm. We compare the inflow generated by live crayfish with that
produced by a mimicking device, a phantom. The phantom consisted of
two nozzles and an inlet providing two jets and a sink so that the
overall mass flux was zero. The use of a phantom enabled us to
analyze the inflow at various jet parameters. We showed that
variation of the jets' directions and relative intensities allows
changing the direction of odour attraction. These results provide a
rationale for biomimetic robot design. We discuss sensitivity and
efficiency of such a robot.}

\vspace{3mm} \noindent Key words: chemical sensing, crayfish, PIV
flow measurement, biomimetics


\vspace{10mm} \centerline{\bf Introduction} \vspace{5mm}

In an aquatic environment, vital information about the presence and
location of food is provided by chemical stimuli. In contrast to
visual and acoustic stimuli, chemicals are dispersed relatively
slowly by molecular diffusion and the ambient flow. Molecular
diffusion is relevant for orientation of microscopic organisms
(Blackburn et al., 1998) being inefficient at providing olfactory
information to animals at distances larger than a few millimetres
(Dusenbery, 1992; Weissburg, 2000). That is why large aquatic
animals such as crayfish rely on the information extracted from
macroscopic odour plumes. These plumes form downstream of the odour
source by advection and turbulent diffusion (Balkovsky and Shraiman,
2002).

Animals can orientate towards an odour source by following the mean
direction of the flow carrying odour molecules (odour-gated
rheotaxis) or by evaluating parameters of the turbulent odour plumes
(Vickers, 2000; Weissburg, 2000). Lobsters, crabs and crayfish have
been shown to navigate towards odour sources using a combination of
these strategies (Atema, 1996; Moore and Grills, 1999; Weissburg and
Derby, 1995; Grasso and Basil, 2002).

Animal orientation under still water conditions such as in ponds,
lakes, caves, or during slack tide, is less studied. In aquatic
environments with little or no ambient water movement, the flow
created by an animal itself could help in odour acquisition and
orientation. Understanding the active olfactory mechanisms of
crayfish inhabiting stagnant waters requires consideration of the
flow patterns created by an animal and the transport of odour
stimulus to the chemoreceptors. This knowledge can be applied to the
design of a robot searching for chemical sources under the stagnant
water conditions.

Crustaceans are well known for their ability to create directed
water currents by pumping and fanning appendages (Herberholz and
Schmitz, 2001; Atema, 1985; Budd et al, 1979; Burrows and Willows,
1969; Brock, 1926). Different appendages can create distinct
currents. For example gill currents ventilate the gills, abdominal
swimmeret currents aid in locomotion, and currents created by the
anterior fan organs have been suggested to be used for odour
acquisition and chemical signalling (Breithaupt, 2001).

\begin{figure}
\centerline{\includegraphics[width=85mm,height=44.6mm]{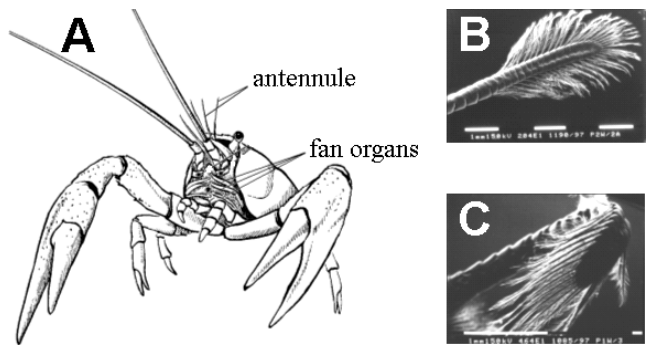}}
\caption{(A) Location of fan organs and major chemoreceptors
(antennules) of crayfish. The fan organs are multisegmental
flagellae (exopodite) of the mouthparts (maxilliped) and are
feathered distally (B). During the power stroke (B; SEM picture) the
feathered hairs are extended. During the recovery stroke (C; SEM
picture) the feathered hairs are folded in. White bars in SEM
pictures are to scale (1 mm). Reprinted from (Breithaupt, 2001) with
permission of Biological Bulletin.} \label{crayfish_fig_appendages}
\end{figure}

The fan organs of a crayfish consist of the flagellae of the three
bilateral maxillipeds. As shown in
Fig.~\ref{crayfish_fig_appendages}A, fan organs are located below
mouth and antennules, which constitute the major chemoreceptors in
decapod crustaceans. The distal part of the multi-segmental flagella
bears a dense single-layered array of feathered hairs emerging on
either side of the stem (Fig. \ref{crayfish_fig_appendages}B).
During a power stroke the hairs are erect, during a recovery stroke
the hairs are tilted and the stem is flexed in order to reduce the
drag (Fig.~\ref{crayfish_fig_appendages}C). The fan organs can be
used for chemical communication by generating a forward directed jet
(Breithaupt and Eger, 2002, Bergman et al., 2005) or for odour
acquisition by drawing water towards the head region (Breithaupt,
2001). In decapod crustaceans, odour stimulation generally initiates
fanning activity (Atema, 1985; Brock, 1926). Crayfish with fan
organs restrained were not able to find the odour source
(Breithaupt, unpublished data).

The red swamp crayfish {\it Procambarus clarkii} inhabits rice
fields and other stagnant water bodies and therefore might utilize
fanning as an active olfactory mechanism. The mechanism by which the
fan-generated currents draw water towards the olfactory appendages
is not understood. Here we investigate how the red swamp crayfish
uses outward directed currents to generate an inflow that could
serve for odour acquisition. We visualized the jets generated by a
crayfish and used Particle Image Velocimetry (PIV) to quantify the
flow velocity around live animal.

In order to test a proposed mechanism for active odour acquisition
we designed a phantom, an assembly of nozzles, simulating the flow
pattern by producing the jets. The phantom enabled us to reproduce
various velocity fields observed in crayfish. Behaviour of
crustaceans has been used as the model for development of autonomous
robots searching for chemical sources (Grasso, 2001; Ayers, 2004;
Ishida et al., 2006; Martinez et al. 2006). Active generation of the
flow might help to locate chemical sources in still fluids. Our
phantom can be used as a prototype design of robots orienting by
chemical clues.

\newpage
\vspace{10mm} \centerline{\bf Materials and methods} \vspace{2mm}

\begin{figure}
\centerline{\includegraphics[width=85mm,height=68.6mm]{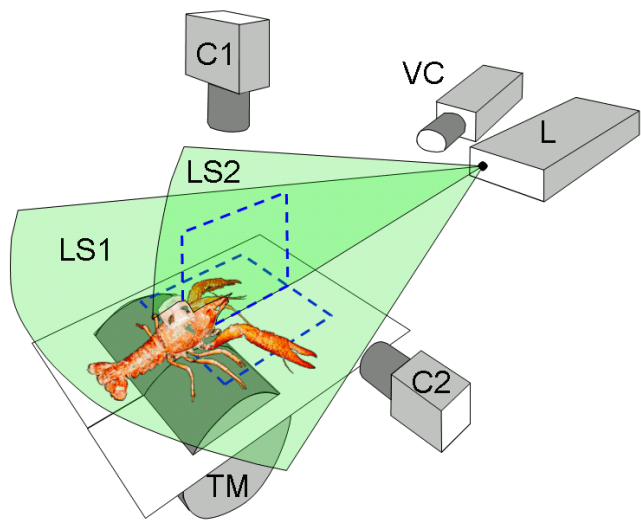}}
\caption{ Experimental setup. A crayfish is suspended above the
treadmill TM. A lasersheet is generated by a laser L. The PIV camera
is set to the position C1(C2) and the lasersheet is aligned along
the plane LS1(LS2) to measure the two components of velocity field
in horizontal(vertical) planes. The video camera VC is used to
monitor the crayfish fanning activity. }\label{crayfish_fig_setup}
\end{figure}

Flow visualization and PIV measurements were conducted for five
adult animals: one female and four males of 7 to 8 cm body length. A
crayfish was suspended with a holder glued to its back in an
aquarium of the size of $60\times60\times120$ cm as shown in Fig.
\ref{crayfish_fig_setup}. A video camera (VC) was used to monitor
activity of fan organs and position of antennules of the crayfish.
The crayfish was suspended above a double wheeled freely rotating
treadmill (TM) with the walking legs of each side resting on a
separate wheel. The treadmill was aimed to reduce agitation of the
animal. It helped the crayfish to rest or to perform natural walking
movements involved in olfactory search behaviour. To prevent animals
from reacting to visual disturbances they were blindfolded by
wrapping an opaque tape around the eyestalks and the rostrum. The
animal claws were removed by inducing autotomy since they interfere
with flow measurements by shadowing the laser-sheet. Experiments
with the unrestrained crayfish showed that claws are kept motionless
during the odour acquisition so that their removal is not likely to
affect the olfactory mechanism (Breithaupt, unpublished data).

We visualized the flow created by a crayfish by introducing
water-based ink in proximity of its fan organs. As it will be
discussed in details, the flow around the crayfish consists of the
outward jets produced by fan organs and the inflow carrying the
odour towards the animal's chemoreceptors. To measure the velocity
field in the inflow area we used a two-dimensional PIV system
(Dantec Dynamics, Ltd) including a 125 mJ double pulse Nd YAG laser
and a Hi-Sense 12-bit double-frame CCD camera (Dantec) with the
resolution 1280$\times$1024 pixels. The FlowMap software (Dantec)
was used to calculate the velocity field with the adaptive
cross-correlation algorithm. Streamlines were calculated using
MATLAB 7.0 software (The MathWorks). Silvered hollow glass spheres
of diameter 10 $\mu$m and density 1.03 g/cm$^3$ were used as seeding
particles. For PIV measurements of the horizontal velocity field the
laser sheet and the camera were aligned at LS1 and C1 positions as
shown in Fig.~\ref{crayfish_fig_setup}. For measurements in the
vertical plane they were set at positions LS2 and C2. Both the laser
source and the camera were fixed on a frame attached to a vertical
translational stage which allows measurement of the velocity field
at various distances from the bottom. A time interval between the
laser pulses was set at 50~ms to measure the low magnitude velocity
in the inflow area. With this time interval the higher velocity
within outward jets could not be resolved. Flow measurements within
the jets were not attempted as the animal often changes the jet
directions and these changes require re-alignment of the laser sheet
and cameras.

After the crayfish was set to the measurement position, it was
allowed to settle down for at least half an hour. After it started
fanning, the PIV was triggered to acquire data at a rate of 1 frame
per second. Only the images acquired during the 30 seconds of each
experimental run were taken for analysis. Following these 30 seconds
the animal often ceased waving its fan organs reacting to the laser
pulses. It is likely that the crayfish retinal or the extra-retinal
(Sandeman et al. 1990, Edwards, 1984) photoreceptors have been
stimulated by the light scattered by the animal's translucent body.

\begin{figure}
\centerline{\includegraphics[width=160mm]{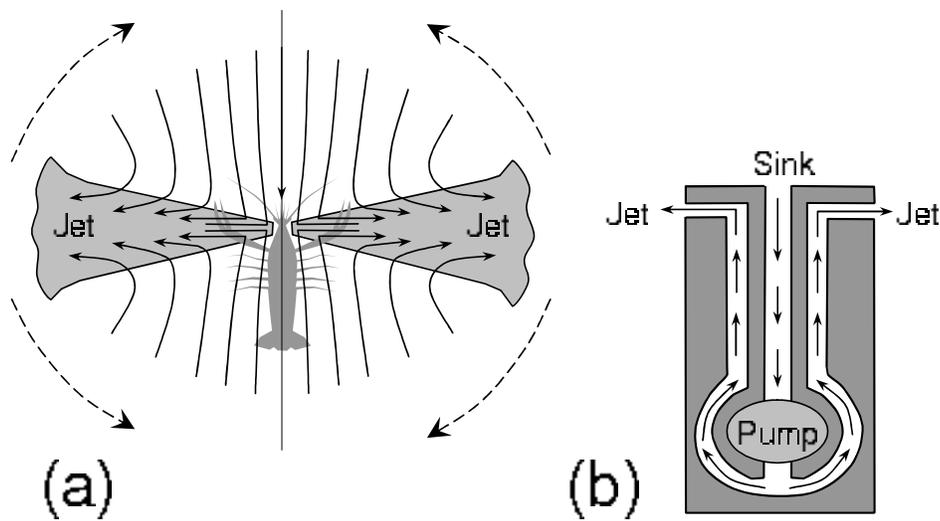}}
\vspace{-80mm}\caption{ The flow generated by crayfish fan organs
consists of the jets and the inflow (a). The inflow  converges
towards the jets' origin and towards jets' axes. At a distance
larger than the jets' length the flow is virtually unaffected by a
crayfish fanning activity. The jets are generally out of horizontal
plane, so the sketch shows a projection. To mimic the flow created
by a crayfish, an assembly of an inlet and two outlet nozzles (b) is
designed. The water is pumped through a closed loop providing the
fluid conservation which is obviously the case for a crayfish. Jets
created by the animal are generally out of horizontal plane and thus
are not presented in the measured flow fields.
}\label{crayfish_fig_model}
\end{figure}

To model the flow generating mechanism involved in chemoreception of
the crayfish we designed a phantom, a closed loop pump-nozzle
assembly with one inlet and two outlets
(Fig.~\ref{crayfish_fig_model}). The phantom simulates the far field
flow of the fan organs which generate water jets preserving the
amount of water involved in the motion. The fluid mass conservation
is achieved by feeding the outlet nozzles with the fluid pumped from
the sink via a closed loop. As in the case with live animals, an
inflow appears to replace the fluid taken in by the sink (and then
ejected as jets), and the fluid entrained by the jets. The phantom
allows to generate horizontal sideward jets, jets directed
45$^\circ$ backwards, and jets directed 45$^\circ$ upwards and is
manufactured in the size of a crayfish. Velocity of the flow induced
by the mimicking device was measured with the same PIV arrangements
used for the crayfish and the flow rate through the nozzles was
adjusted to make the magnitude of the inflow close to that observed
in experiments with live animals.

\vspace{10mm} \centerline{\bf Results} \vspace{5mm}

\begin{figure}
\centerline{\includegraphics[width=100mm]{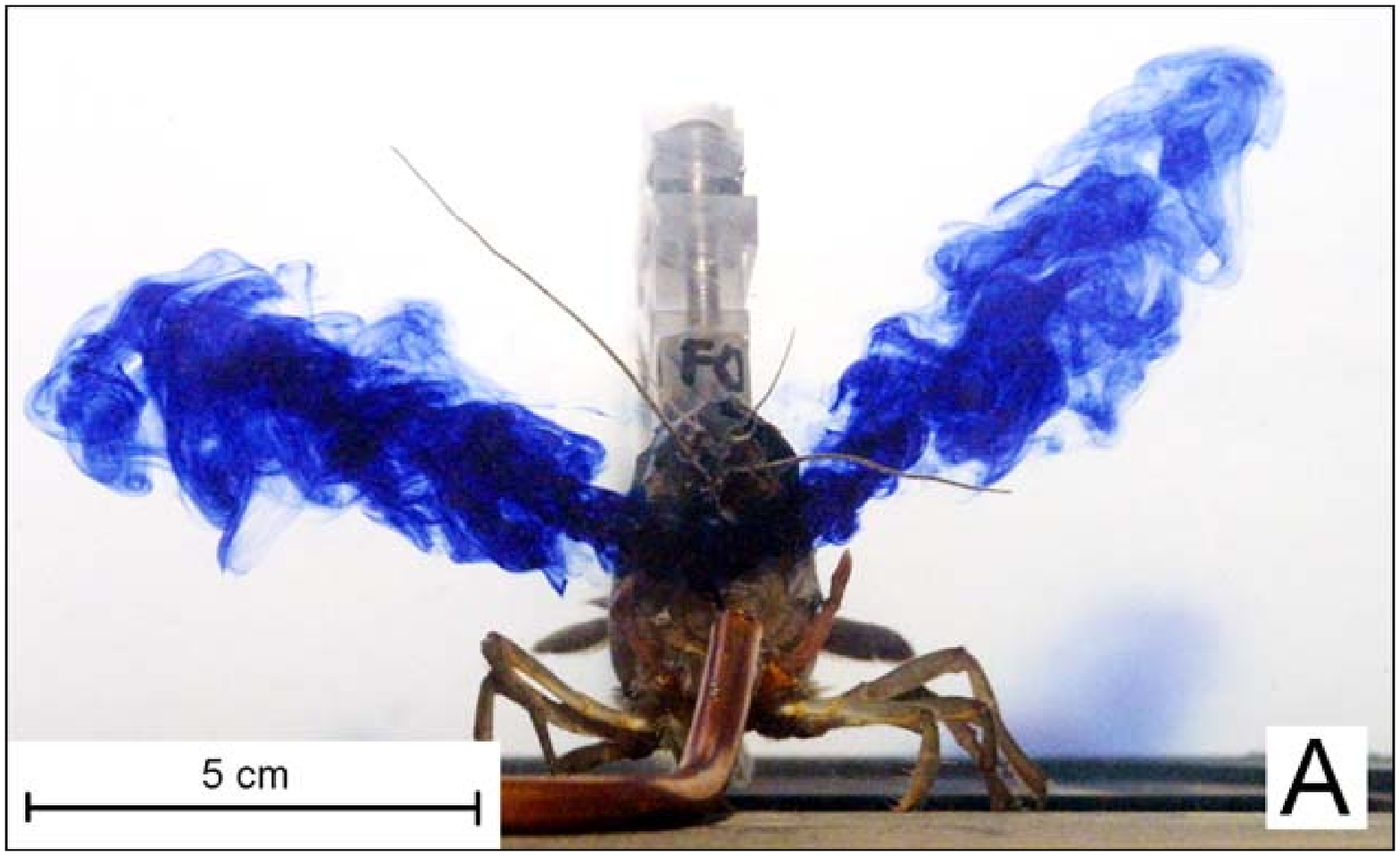}}\break\break
\centerline{\includegraphics[width=100mm]{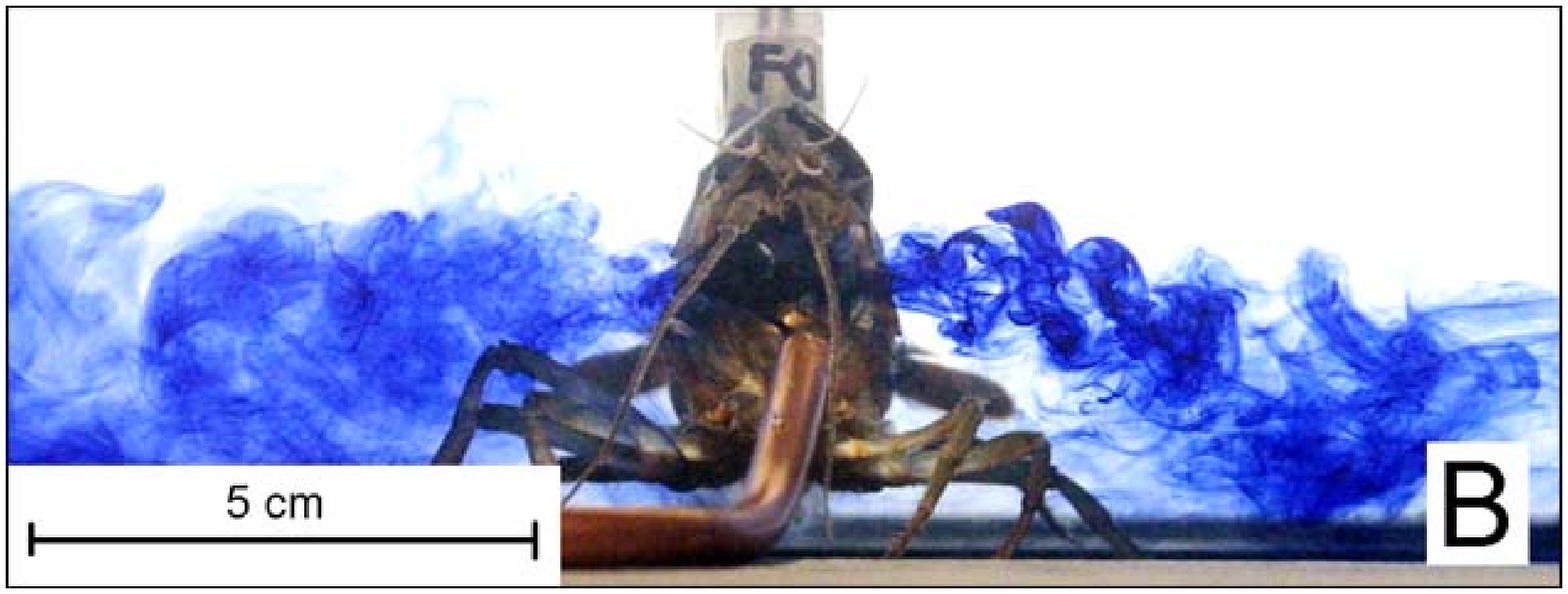}}\break\break
\centerline{\includegraphics[width=100mm]{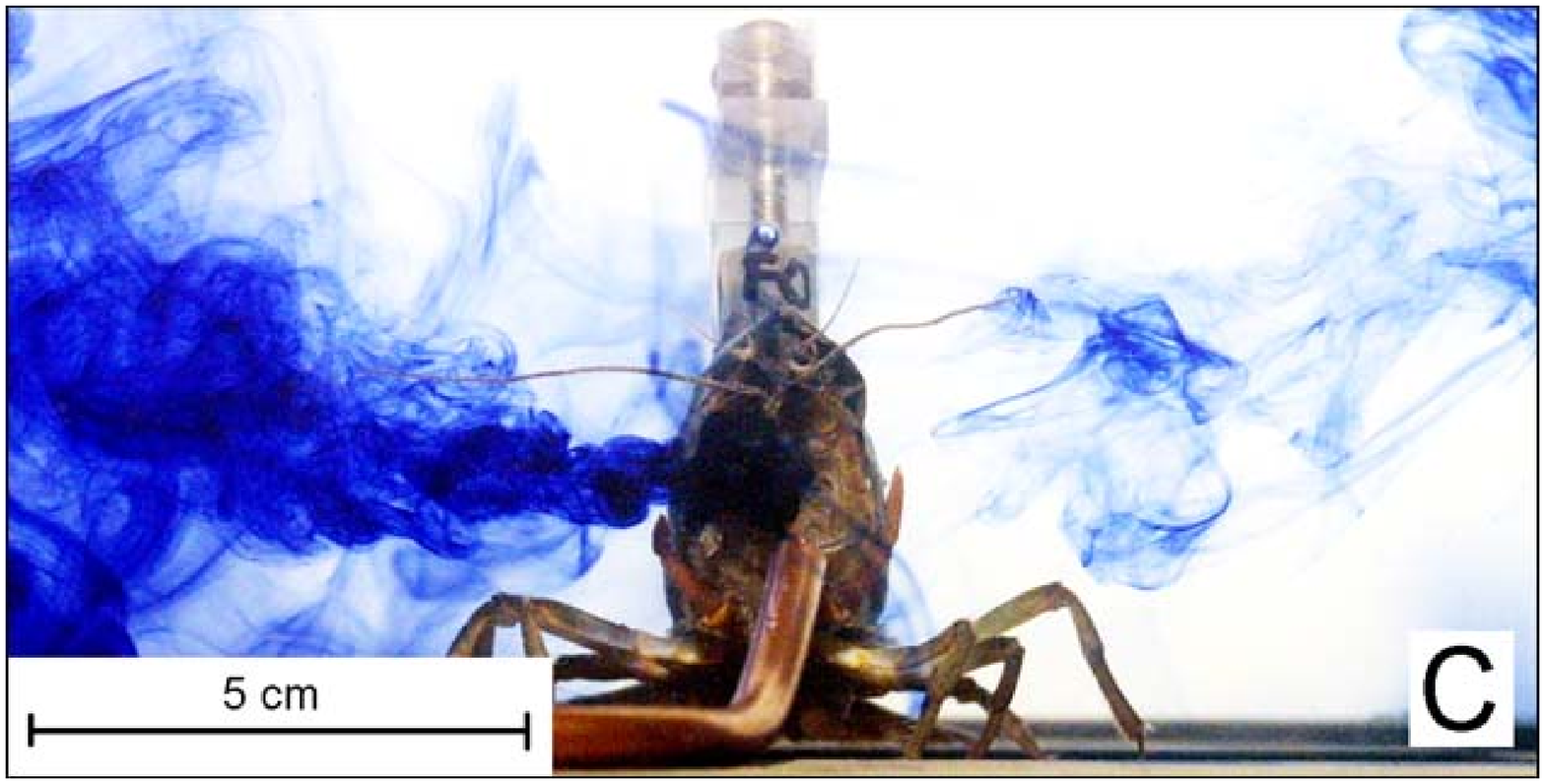}}
\caption{Ink visualization of the jets generated by the crayfish fan
organs. Bilateral jets directed at $25^\circ$ and $40^\circ$ upwards
(a) and horizontally (b); unilateral jet directed horizontally (c).
The ink is slowly released from a pipe in front of the animal fan
organs. }\label{crayfish_fig_ink}
\end{figure}

A crayfish produced one or two jets by waving the fan organs of one
or of either sides (Fig. \ref{crayfish_fig_ink}). Jets' direction
varied from 90$^\circ$ to the animal plane of symmetry to 45$^\circ$
upwards and 45$^\circ$ backwards. The jets of the length more than
10~cm were observed. Fanning was sometimes accompanied by flicking
the antennules, a behaviour that enhances olfaction in decapod
crustaceans (Schmitt and Ache, 1979). The animals may occasionally
walk on the treadmill while waving the fan organs, but most of the
time their walking legs are steady.

\begin{figure}
\centerline{\includegraphics[width=80mm]{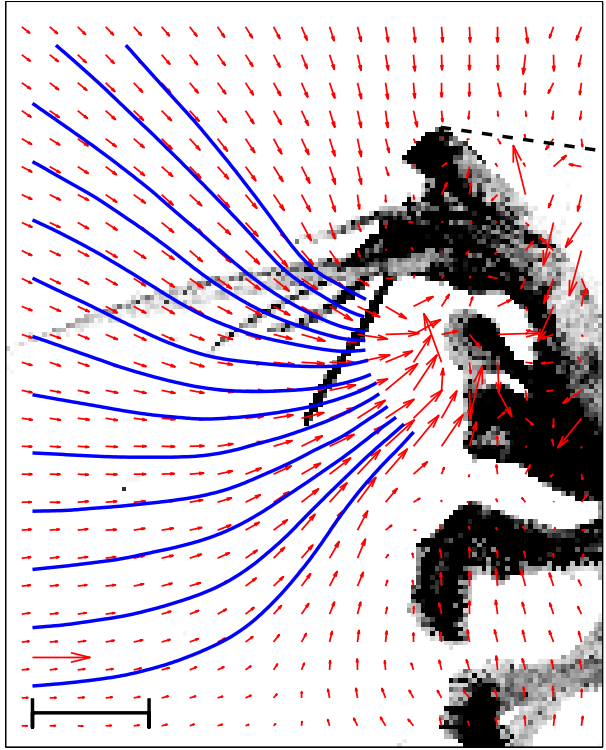}}
\caption{The flow field measured at the crayfish plane of symmetry
(lasersheet LS2 in Fig.~\ref{crayfish_fig_setup}). Average over 30
instantaneous measurements (30 seconds). A negative image of the
crayfish cut from a PIV image is shown for reference. Observe that
antennules with the chemoreceptors are lowered in front of fan
organs. A reference vector at the bottom left is 1~cm/s, a reference
segment is 1~cm.}\label{crayfish_fig_PIV_animal_vert}
\end{figure}

\begin{figure}
\includegraphics[width=135mm]{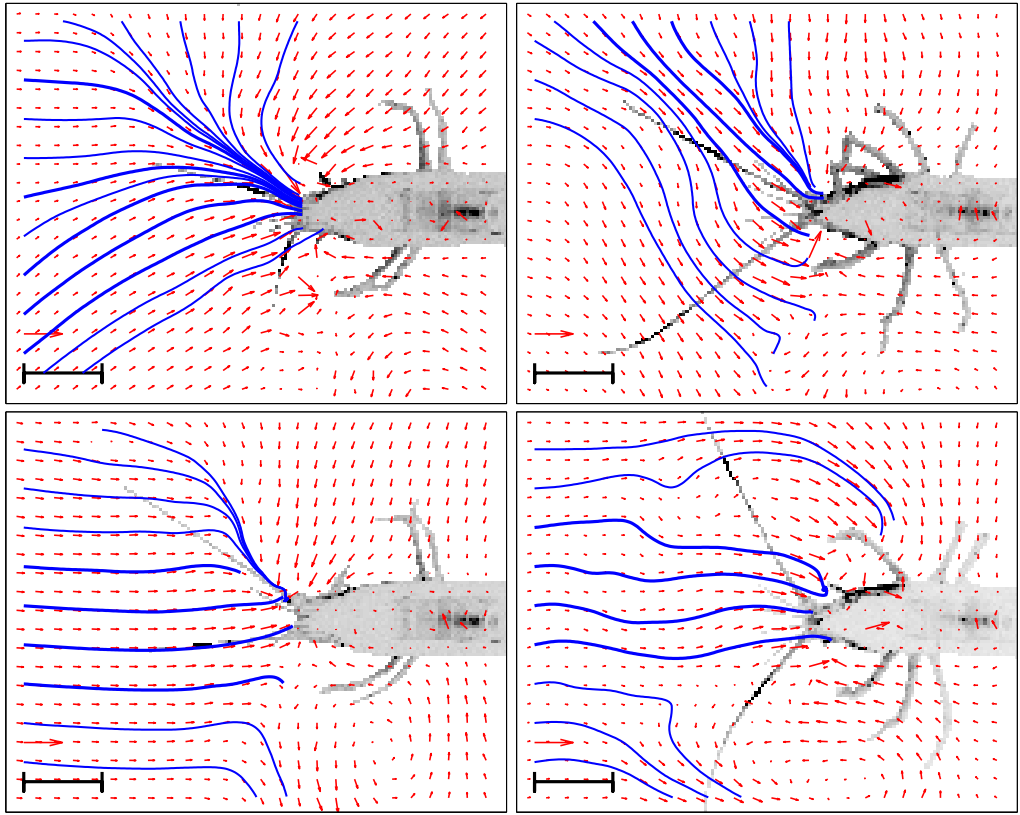}
\caption{ Velocity field generated by a crayfish measured in a
horizontal plane LS1 (Fig.~\ref{crayfish_fig_setup}). Average over
30 instantaneous measurements (30 seconds). A negative image of the
animal cut from a PIV image is shown for a reference. A reference
vector at the bottom left is 1~cm/s, a reference segment is 2~cm.
The jets created by the animal are outside the plane of measurement
and thus are not observed in the vector field.
}\label{crayfish_fig_PIV_animal}
\end{figure}

\begin{figure}
\includegraphics[width=135mm]{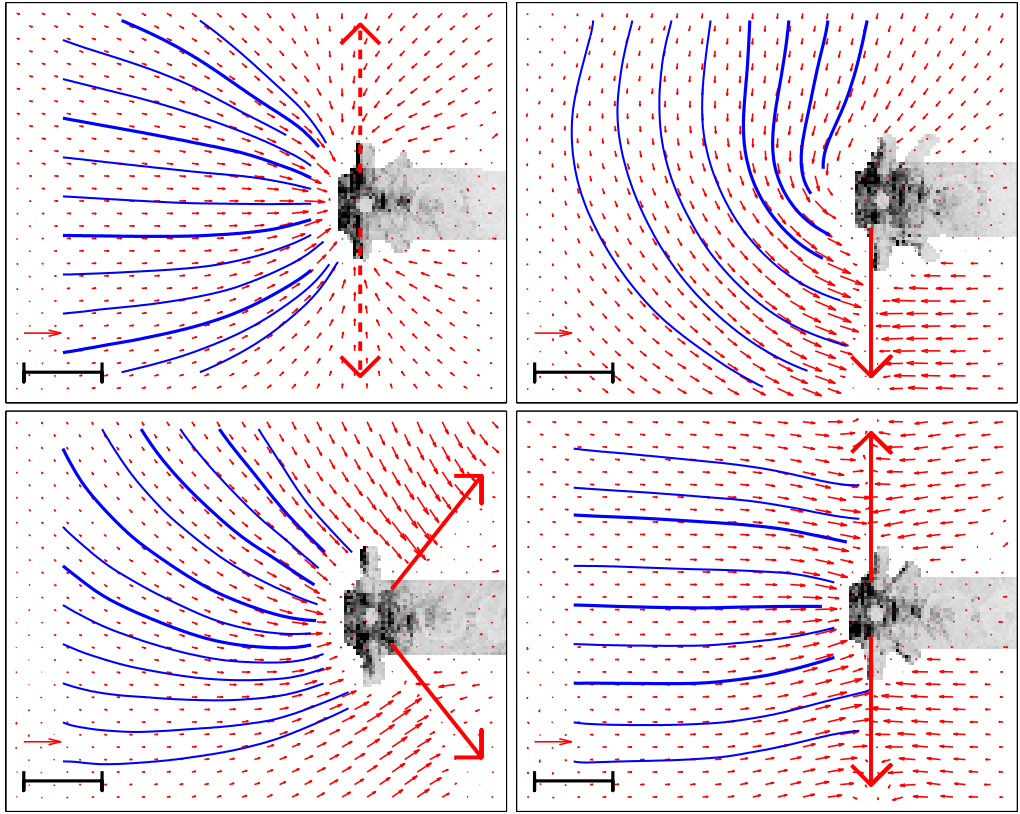}
\caption{Velocity field generated by a mimicking device
(Fig.~\ref{crayfish_fig_model}b) measured in a horizontal plane LS1
(Fig.~\ref{crayfish_fig_setup}). A negative image of the device cut
from a PIV image is shown for a reference. Positions of the jets
produced by a model are indicated by bold lines. Two jets directed
45$^\circ$ upwards, off-plane (a), a single jet directed to the left
(b), the two jets directed 45$^\circ$ backwards (c), and two jets
directed to the sides (d). A reference vector at the bottom left is
1~cm/s, a reference segment is 2~cm. Short arrows at jet locations
and over the model correspond to the 'noise' of image processing
software and are not indicating any real values of fluid
velocity.}\label{crayfish_fig_PIV_model}
\end{figure}

The outward jets induce an inflow converging towards the fan organs
and the jets themselves. A typical flow field measured by PIV in the
vertical plain is shown in Fig.~\ref{crayfish_fig_PIV_animal_vert}.
The antennules are lowered in front of the fan organs thereby
exposing olfactory receptors to the incoming flow. The streamlines
in Fig.~\ref{crayfish_fig_PIV_animal_vert} are more or less
horizontal at the level of fan organs which suggests that one can
study the flow structure around the animal by measuring the flow
field at this hight. Four examples of the flow field measured in the
horizontal plane are shown in Fig.~\ref{crayfish_fig_PIV_animal}.
The streamlines leading to the animal chemoreceptors illustrate that
the sector of odour attraction may vary. The jets are out of the
plane of measurement in most of experiments with live animals, which
may create an illusion of breaking the fluid mass conservation. Four
samples of the flow field generated by the mimicking device are
shown in Fig.~\ref{crayfish_fig_PIV_model}. The inflow patterns
illustrate how the flow outside the jets converges towards the jets'
origin (nozzles), and towards the jet axis.

A reasonable question to ask is how long does it take for the inflow
pattern to adjust to the changed configuration of the jets. Since
the animal's fanning behaviour is out of our direct control, we have
performed experiments with the phantom. Several experimental runs
with the flow through nozzles switched on from the rest showed that
the steady flow field is set within several seconds after starting
the pump.

\begin{figure}
\includegraphics[width=135mm]{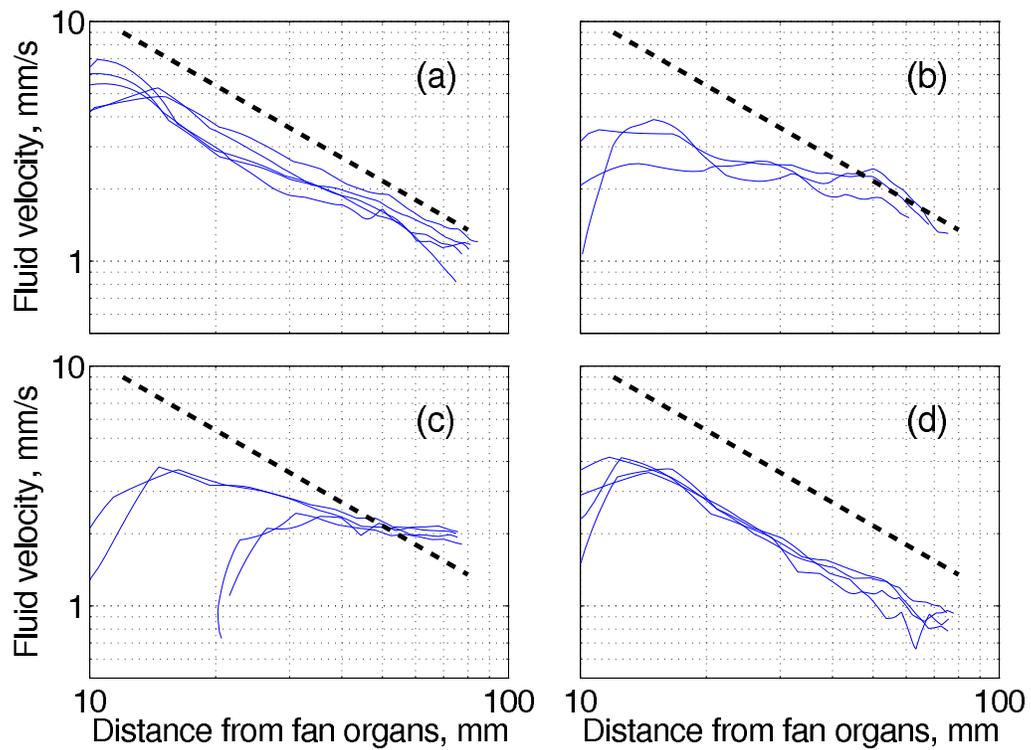}
\caption{Absolute value of fluid velocity plotted $vs.$ distance
from the crayfish fan organs along the streamlines shown in bold in
Fig.~\ref{crayfish_fig_PIV_animal}. The dashed lines correspond to
$V\propto1/s$ as in formula (\ref{crayfish_V}).
}\label{crayfish_fig_veloprofiles_animal}
\end{figure}

\begin{figure}
\includegraphics[width=135mm]{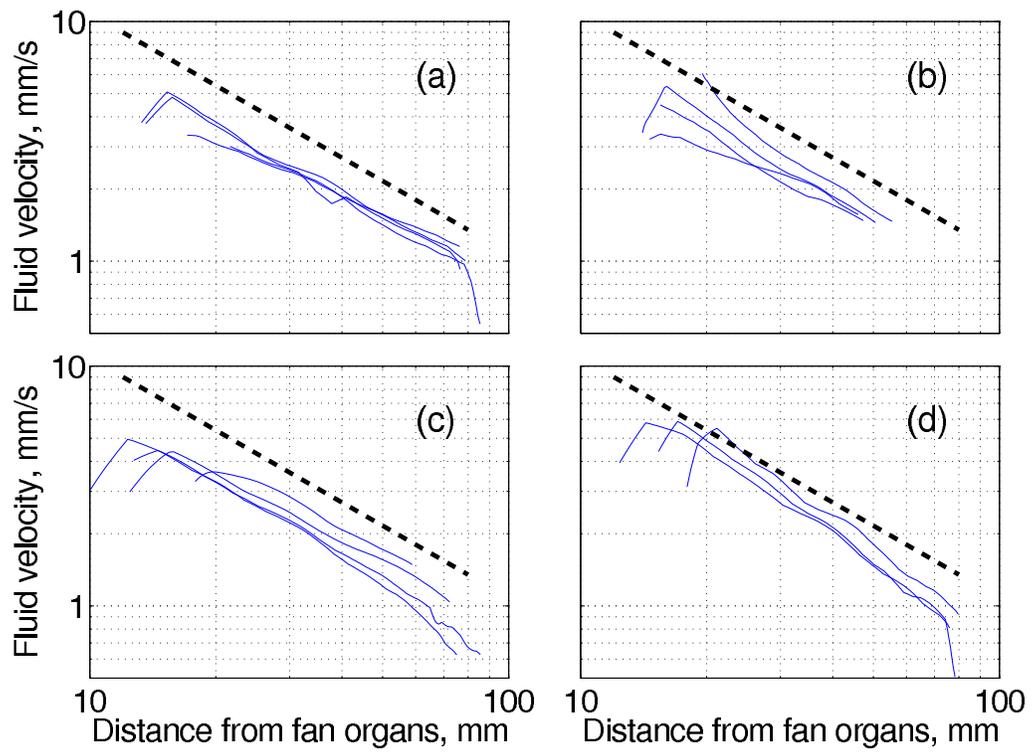}
\caption{Absolute value of fluid velocity $vs.$ distance from the
jets origin along the streamlines shown in bold in
Fig.~\ref{crayfish_fig_PIV_model}. The dashed line line corresponds
to $V\propto1/s$ as in formula
(\ref{crayfish_V}).}\label{crayfish_fig_veloprofiles_model}
\end{figure}

The flow velocity as a function of distance from the fan organs/sink
along the streamlines is plotted in Figs.
\ref{crayfish_fig_veloprofiles_animal} and
\ref{crayfish_fig_veloprofiles_model}. The slopes of the plots show
that the flow velocity is somewhat inversely proportional to the
distance from the fan organs i.e.
\begin{eqnarray}
 && V\sim V_0\frac{s_0}{s}\label{crayfish_V}
\end{eqnarray}
where $V_0$ is the fluid velocity at a distance $s_0$ from the fan
organs. The time required for the odour patch located at a distance
$L$ from the fan organs to reach the crayfish antennules can be
estimated by integrating inverse velocity by the distance:
\begin{eqnarray}
 && T\sim \int_0^L \frac1V~ds=\int_0^L \frac{1}{V_0}\frac{s}{s_0}~ds=\frac{L^2}{2V_0s_0}\label{crayfish_T}
\end{eqnarray}
Assuming the velocity $V_0\approx5$~mm/s at $s_0=10$~mm (as in
Fig.~\ref{crayfish_fig_veloprofiles_animal}) we can infer that the
odour patch from stimulus located at 100~mm from the crayfish fan
organs would reach the chemoreceptors in approximately 100~s. This
time interval increases quadratically with $L$, rising to
approximately 4~min for the distance of 150~mm. For comparison,
solution of the diffusion equation for the point source (Batchelor,
1970, p.187) shows that the characteristic time of molecular
diffusion at a distance of 100~mm is measured in days.

\vspace{10mm} \centerline{\bf Discussion} \vspace{5mm}

We have described a mechanism which crayfish may utilize to assess
odour stimuli in a stagnant water environment. Using the anterior
fan organs a crayfish creates jets (Fig.~\ref{crayfish_fig_ink}) so
that the flow induced by these jets
(Figs.~\ref{crayfish_fig_PIV_animal_vert},~\ref{crayfish_fig_PIV_animal})
draws a water sample from a distance to the animal's chemoreceptors.
Schematic of the flow around the crayfish is sketched in
Fig.~\ref{crayfish_fig_model}.

The most important fact revealed by the PIV measurements is a
surprisingly slow decrease of the inflow velocity with the distance
from fan organs, the jets' origin
(Figs.~\ref{crayfish_fig_veloprofiles_animal},~\ref{crayfish_fig_veloprofiles_model}).
The slow decay is explained by the fluid entrainment by jets. If the
entrainment was insignificant and the inflow was formed only by a
point sink at the location of fan organs (the jets' origin), one
would expect the inflow to be spherically symmetric (Batchelor,
1970, p.89). In that case the flow velocity would decrease as the
inverse square of the distance from fan organs, and the travel time
$T$ for the odour patch to reach the crayfish chemoreceptors
(equation \ref{crayfish_T}) would be proportional to the 3rd power
of the distance $L$ to the odour source. This would increase the
travel time to 10~min for a distance of 100~mm and to 40~min for a
distance of 150~mm. Such time intervals are beyond the biologically
relevant time scale. Note that, despite slow decrease with the
distance, the inflow induced by jets is limited in space by the
length of these jets as illustrated in
Fig.~\ref{crayfish_fig_model}a.

Fluid entrainment by a single jet has been described theoretically
by Schneider (1981). He showed that a turbulent jet acts as a line
sink with respect to the surrounding fluid. This leads to the axial
rather than spherical symmetry of the flow field, hence to a slower
decay of velocity with the distance. The decay is now proportional
to inverse distance from the jet axis (a line) instead of inverse
squared distance from the sink (a point). The secondary flow induced
by a turbulent jet and decay of the jet itself were described
analytically by Kotsovinos and Angelidis (1991). However, results of
this paper can not be extended to a case of two jets because the
Navier-Stokes equations governing fluid motion are not linear.
Moreover, the relatively weak jets created by a crayfish can not be
considered as fully developed turbulent jets, and presence of a
rigid boundary, the bottom of the tank, affects the flow. The above
argument suggests employment of direct numerical simulation to model
the flow induced by the jets, and the numerical simulation of a
system of jets is a problem on its own. In this work, we do not go
further in the investigation of a particular inflow structure,
observing that the range of odour acquisition, i.e. the range where
the inflow rate decreases slowly with the distance, is now defined
by the jets' length which may exceed 10~cm.

To locate a source of the odour the crayfish needs to define a
direction to this source. In a riverine environment the source is
somewhere upstream. In still water an animal can navigate comparing
the local intensity of the odour at different locations through the
plume (Atema 1996). However, the animal movement may stir up the
water and destroy the pattern of odour patches formed by the source.
Active movement would also increase predation risks and may be
energetically expensive. The crayfish can overcome these
difficulties by scanning the environment employing a selection of
jet patterns intended to provide varying directions of incoming flow
as shown in
Fig.~\ref{crayfish_fig_PIV_animal},~\ref{crayfish_fig_PIV_model}.
Predation risks are now significantly reduced since hydrodynamic
disturbance created by jets is detectable from a shorter distance in
comparison with the moving visual target. When the bottom of a pond
or a lake is covered by plants, scanning of environment with the
help of jets becomes also energetically profitable since the stalks
are less an obstacle to the flow than they are to the walking
crayfish.

To show that production of the jets is not energetically expensive,
we make an estimate based on parameters measured by Breithaupt
(2001). Consider 6 appendages of the area $A=10\,$mm$^2$ each waving
with the frequency $f=6\,$Hz, amplitude $a=5\,$mm, and velocity
$u=2\pi f a\approx 20\,$cm/s. Water density is
$\rho=1000\,$kg/m$^3$, its kinematic viscosity is
$\nu=0.01\,$cm$^2$/s. The Reynolds number of the flow around an
appendage can be estimated as
\begin{eqnarray}
&&{\rm Re}=\frac{u \cdot a}{\nu}=\frac{20\,{\rm cm\,s}^{-1}\cdot
0.5\,{\rm cm}}{0.01\,{\rm cm}^2\,{\rm s}^{-1}}\approx 1000,
\end{eqnarray}
so that we can assume that the resistance force overcame by an
appendage is defined by the dynamic term
\begin{eqnarray}
&& F=\frac12\rho u^2\cdot A.
\end{eqnarray}
The energy required to wave the 6 appendages for a week is
\begin{eqnarray}
 E\approx 604800\frac{\rm s}{\rm week}\cdot f\cdot 6\,{\rm appendages}\cdot F\cdot a \approx 20\,{\rm J}\approx 5\,{\rm Calories}
\end{eqnarray}
Allowing for the variability of the appendage area, the beat
frequency, and the efficiency of the muscles driving the fan organs
(10\%) the estimated energy can rise to 200\,Calories which is still
small compared to the total energy gained by feeding in a week.

Observations of an unrestrained crayfish show that it may fan the
appendages both when walking and when hiding in a shelter. This
suggests that the animal creates the inflow both to assist active
search for food and to detect the food appearance in the vicinity of
the shelter. To understand the exact way in which the crayfish
utilizes its self-generated inflow, behavioural experiments are
required with unrestrained animals in a large tank.

Experiments with the phantom (Fig.~\ref{crayfish_fig_PIV_model})
helped to understand the mechanism of odour attraction employed by
the crayfish. This mechanism could well be adapted to design better
robots for finding chemical sources in stagnant fluids. Nozzle
assemblies generating one or several jets should be more efficient
in drawing distant odour patches to the sensor than devices that
create a sink flow. Unlike the live animals, a man-made mechanic
assembly has fewer restrictions on the parameters affecting
performance of the system. By optimizing nozzle profiles the jets
can be made more regular and by increasing the pump power jets can
be made considerably longer than those generated by a crayfish (up
to metres). A number of jets and their alignment can be altered to
obtain a desirable pattern of the inflow. Placing a set of sensors
at different locations around the nozzles would enable measurement
of the direction where the odour is coming from. Measurements of the
time lag between switching the jets on and arrival of a chemical
would add the data about the distance to the source.

The range of odour acquisition (the range of a relatively slow
decrease of the inflow rate with the distance from the generating
assembly) and the directional sensitivity of the robot is defined by
the length of jets rather than by physical size of the device. This
would allow design of a relatively small device with an ability to
access the obstructed area of interest. Moreover, induction of the
inflow by the outward jets would eliminate necessity of direct
access to the area enabling, for example, chemical sensing through
the mesh. Devices similar to that described above can be employed to
search leaks from the pipes, to search chemical sources at the ocean
bed, or for the non-invasive flow monitoring. The idea of using a
jet to create an inflow is not restricted to the aquatic environment
and can be used, for example, for the non-invasive monitoring of the
agricultural land.

We would like to acknowledge financial support of this study by the
Hull Environmental Research Institute to T.B. and S.L. and a NERC
fellowship NER/I/S/2000/01411 to T.B.

\vspace{10mm} \centerline{\bf References} \vspace{5mm}


\noindent 
{Atema, J.} (1996) Eddy chemotaxis and odor landscapes: exploration
of nature with animal sensors. {\it Biol. Bull.} {\bf 191},
129--138. \vspace{1mm}

\noindent 
{Atema, J.} (1985) Chemoreception in the sea: adaptations of
chemoreceptors and behaviour to aquatic stimulus conditions. {\it
Symp. Soc. Exp. Biol.} {\bf 39} (1985) 386--423. \vspace{1mm}

\noindent 
{Ayers, J.} (2004) Underwater walking. {\it Arthropod. Struct. Dev.}
{\bf 33}, 347--360. \vspace{1mm}

\noindent 
{Balkovsky, E., Shraiman, B.I.} (2000) Olfactory search at high
Reynolds number. {\it PNAS} {\bf 99(20)}, 12589--12593. \vspace{1mm}

\noindent 
{Batchelor, G.K.} (1970) An introduction to Fluid Dynamics. {\it
Cambridge University Press}.\vspace{1mm}

\noindent 
{Bergman, D.A., Martin, A.L. and Moore, P.A.} (2005) Control of
information flow through the influence of mechanical and chemical
signals during agonistic encounters by the crayfish, {\it Orconectes
rusticus}. {\it Animal Behaviour} {\bf 70}, 485--496.\vspace{1mm}

\noindent 
{Blackburn, N., Fenchel, T., and Mitchell, J.} (1998) Microscale
nutrient patches in planktonic habitats shown by chemotactic
bacteria. {\it Science} {\bf 282}, 2254--2256. \vspace{1mm}

\noindent 
{Breithaupt, T.} (2001) Fan organs of crayfish enhance chemical
information flow. {\it Biol. Bull.} {\bf 200}, 150--154.
\vspace{1mm}

\noindent 
{Breithaupt, T. and Eger, P.} (2002) Urine makes the difference:
chemical communication in fighting crayfish made visible. {\it J.
Exp. Biol.} {\bf 205}, 1221--1231. \vspace{1mm}

\noindent 
{Brock, F.} (1926) Das Verhalten des Einsiedlerkrebses Pagurus
arrosor Herbst wahrend der Suche und Aufnahme der Nahrung. {\it
Zeitschrift fur Morphologie und Okologie der Tiere} {\bf 6},
415--552. \vspace{1mm}

\noindent 
{Budd, T.W., Lewis, J.C. and Tracey, M.L.} (1979) Filtration feeding
in Orconectes propinquus and Cambarus robustus (Decapoda,
Cambaridae). {\it Crustaceana Supplement} {\bf 5}, 131--134.
\vspace{1mm}

\noindent 
{Burrows, M. and Willows, A.O.D.} (1969) Neuronal co-ordination of
rhythmic maxilliped beating in brachyuran and anomuran crustacea.
{\it Compar. Biochem. Phys.} {\bf 31}, 121--135. \vspace{1mm}

\noindent 
{Dusenbery, D.B.} (1992) {\it Sensory Ecology.} {New York:
W.H.Freeman and Co.} \vspace{1mm}

\noindent 
{Edwards, D.H., Jr.} (1984) Crayfish extraretinal photoreception. I.
Behavioral and motorneuronal responses to abdominal illumination.
{\it J. Exp. Biol.} {\bf 109}, 291--306. \vspace{1mm}

\noindent 
{Grasso, F.W.} (2001) Invertebrate-inspired sensory-motor systems
and autonomous, olfactory-guided exploration. {\it Biol. Bull.} {\bf
200} 160--168. \vspace{1mm}

\noindent 
{Grasso, F.W., Basil, J.A.} (2002) How lobsters, crayfishes, and
crabs locate sources of odor: current perspectives and future
directions. {\it Curr. Opin. Neurobiol.} {\bf 12}, 721--727.
\vspace{1mm}

\noindent 
{Herberholz, J. and Schmitz, B.} (2001) Signaling via Water Currents
in Behavioral Interactions of Snapping Shrimp (Alpheus
heterochaelis). {\it Biol. Bull.} {\bf 201}, 6--16. \vspace{1mm}

\noindent 
{Ishida, H., Tanaka, H., Taniguchi, H., Moriizumi, T.} (2006) Mobile
robot navigation using vision and olfaction to search for a gas/odor
source. {\it Auton. Robot.} {\bf 20}, 231--238. \vspace{1mm}

\noindent 
{Kotsovinos, N.E. and Angelidis, P.B.} (1991) The momentum flux in
turbulent submerged jets. {\it J. Fluid Mech.} {\bf 229}, 453-470.
\vspace{1mm}

\noindent 
{Martinez, D., Rochel, O., Hugues, E.} (2006) A biomimetic robot for
tracking specific odors in turbulent plumes. {\it Auton. Robot} {\bf
20}, 185--195. \vspace{1mm}

\noindent 
{Moore, P.A. and Grills, J.L.} (1999) Chemical orientation to food
by the crayfish Orconectes rusticus: influence of hydrodynamics.
{\it Animal Behaviour} {\bf 58}, 953--963. \vspace{1mm}

\noindent 
{Sandeman, D.C., Sandeman, R.E., \& de Couet, H.G.} (1990)
Extraretinal photoreceptors in the brain of the crayfish Cherax
destructor. {\it J. Neurobiol.} {\bf 21}, 619--29.\vspace{1mm}

\noindent 
{Schmitt, B.C. \& Ache, B.W.} (1979) Olfaction: Responses of a
decapod crustacean are enhanced by flicking. {\it Science}, {\bf
205}, 204--206. \vspace{1mm}

\noindent 
{Schneider, W} (1981) Flow induced by jets and plumes. {\it J. Fluid
Mech.} {\bf 108}, 55--65. \vspace{1mm}

\noindent 
{Vickers, N.J.} (2000) Mechanisms of animal navigation in odor
plumes. {\it Biol. Bull.} {\bf 198}, (2000) 203--212. \vspace{1mm}

\noindent 
{Weissburg, M.J.} (2000) The Fluid Dynamical Context of Chemosensory
Behavior. {\it Biol. Bull.} {\bf 198}, 188--202. \vspace{1mm}

\noindent 
{Weissburg, M.J. and Derby, C.D.} (1995) Regulation of sex-specific
feeding behavior in fiddler crabs: physiological properties of
chemoreceptor neurons in claws and legs of males and females. {\it
J. Compar. Phys. A} {\bf 176}, 513--526. \vspace{1mm}


\vfil\newpage


\end{document}